\def\etbr{$\kappa$-(BE\-DT\--TTF)$_2$\-Cu\-[N\-(CN)$_{2}$]Br}
\def\etcl{$\kappa$-(BE\-DT\--TTF)$_2$\-Cu\-[N\-(CN)$_{2}$]Cl}

\def\etcn{$\kappa$-(BE\-DT\--TTF)$_2$\-Cu$_2$(CN)$_{3}$}
\def\kbr{$\kappa$-Br}
\def\kcl{$\kappa$-Cl}
\def\kcn{$\kappa$-CN}
\def\cm{cm$^{-1}$}
\documentclass[aps,amsmath,amssymb,prb,superscriptaddress,twocolumn,showpacs]{revtex4}
\usepackage{graphicx}
\begin{document} 
\title{Absence of charge order in the dimerized $\kappa$-phase BEDT-TTF salts}
\author{Katrin Sedlmeier}
\affiliation{1.~Physikalisches Institut, Universit\"{a}t
Stuttgart, Pfaffenwaldring 57, D-70550 Stuttgart Germany}
\author{Sebastian Els\"asser}
\affiliation{1.~Physikalisches Institut, Universit\"{a}t
Stuttgart, Pfaffenwaldring 57, D-70550 Stuttgart Germany}
\author{Rebecca Beyer}
\affiliation{1.~Physikalisches Institut, Universit\"{a}t
Stuttgart, Pfaffenwaldring 57, D-70550 Stuttgart Germany}
\author{Dan Wu}
\affiliation{1.~Physikalisches Institut, Universit\"{a}t
Stuttgart, Pfaffenwaldring 57, D-70550 Stuttgart Germany}
\author{Tomislav Ivek}
\affiliation{1.~Physikalisches Institut, Universit\"{a}t
Stuttgart, Pfaffenwaldring 57, D-70550 Stuttgart Germany}
\affiliation{Institut za fiziku, P.O.\ Box 304, HR-10001 Zagreb, Croatia}
\author{Silvia Tomi\'{c}}
\affiliation{Institut za fiziku, P.O.\ Box 304, HR-10001 Zagreb, Croatia}
\author{John A. Schlueter}
\affiliation{Material Science Division, Argonne National Laboratory,
Argonne, Illinois 60439-4831, U.S.A.}
\author{Martin Dressel}
\affiliation{1.~Physikalisches Institut, Universit\"{a}t
Stuttgart, Pfaffenwaldring 57, D-70550 Stuttgart Germany}
\date{\today}
\begin{abstract}
Utilizing infrared vibrational spectroscopy we have investigated
dimerized two-dimensional organic salts in order to search for
possible charge redistribution that might cause electronic dipoles
and ferroelectricity: the quantum spin liquid
$\kappa$-(BE\-DT\--TTF)$_2$\-Cu$_2$(CN)$_{3}$, the
antiferromagnetic Mott insulator
$\kappa$-(BE\-DT\--TTF)$_2$\-Cu\-[N\-(CN)$_{2}$]Cl, and the
superconductor $\kappa$-(BE\-DT\--TTF)$_2$\-Cu\-[N\-(CN)$_{2}$]Br.
None of them exhibit any indication of charge disproportionation
upon cooling down to low temperatures. No
modification in the charge distribution is observed around $T=6$~K
where a low-temperature anomaly has been reported for the spin-liquid
$\kappa$-(BE\-DT\--TTF)$_2$\-Cu$_2$(CN)$_{3}$. In this compound
the in-plane optical response and vibrational coupling
are rather anisotropic, indicating that the tilt of the BEDT-TTF molecules
in $c$-direction and their coupling to the anion layers has to be considered in the explanation of the electromagnetic properties.
\end{abstract}

\pacs{
71.30.+h, 
75.10.Kt  
74.70.Kn,  
78.30.Jw    
}

\maketitle
%
%

\section{Introduction}
Electronic ferroelectricity and magneto-dielectric phenomena are
one of the central issues in solid state physics in recent years
due to fundamental questions as well as possible applications. It
is extensively explored in inorganic transition metal oxides such
as LuFe$_2$O$_4$ or YbFe$_2$O$_4$ \cite{Ikeda05,Ishihara10} as
well as organic charge-transfer salts
(TMTTF)$_2X$.\cite{Monceau01,Brazovskii08} An additional aspect is
brought in by multiferroics, i.e.\ materials with coupled
electric, magnetic, and structural order parameters that result in
simultaneous ferro\-electricity, ferro\-magnetism, and
ferro\-elasticity.\cite{Wang03,Kimura03} While there are different
scenarios realized by now,\cite{Wang09,Tokura10} here we focus on
multi\-ferroicity due to charge ordering, described in the seminal
review by van den Brink and Khomskii.\cite{Brink08} In
charge-driven ferroelectricity a large magneto-dielectric coupling
and a fast polarization switching are expected, since the electric
polarization is governed by electrons compared to ions in
conventional ferro\-electrics.\cite{LinesGlass77}

These phenomena are well studied in one-dimen\-sion\-al organic
TMTTF salts%
,\cite{Pouget96,Dressel07,Brazovskii07,Pouget12} for which charge
order was proven by NMR
spectroscopy,\cite{Chow00,Zamborszky02,Yu04,Fujiyama04} dielectric
permittivity measurements,\cite{Nad00,Monceau01,Nad02,Nad06,Langlois10} Raman
and infrared investigations,\cite{Dumm04,Furukawa05,Dressel12} thermal expansion,\cite{deSouza08} and neutron scattering experiments.\cite{Foury10}
Calculations based on the extended Hubbard Hamiltonian could
reproduce the charge-ordered state.\cite{Seo04,Clay03} The charge
per molecule alternates along the stacking direction $a$ between
charge rich ($\rho_0+\delta$) and charge poor molecules
($\rho_0-\delta$). Recent ESR experiments strongly suggest that
two inequivalent magnetic TMTTF chains coexist in the
charge-ordered regime,\cite{Yasin12} most probably due to the loss
of translational invariance in the $bc$-plane. The increase of
mosaicity around $T_{\rm CO}$ inferred by $X$-ray investigations
supports the development of ferroelectric domains in the nanometer
scale.\cite{Rose12}

As far as organic crystalline materials are concerned, the
two-dimensional BEDT-TTF salts with quarter-filled conduction
band, such as $\alpha$-(BEDT-TTF$_2$I$_3$ and
$\theta$-(BEDT-TTF)$_2$RbZn(SCN)$_4$, have been established as
model compounds of charge order for theory \cite{Seo04} and
experiment.\cite{Takahashi06} The possibility of
charge-disproportionation in dimerized half-filled BEDT-TTF
systems, is intensely debated for the last years,
\cite{Hotta10,Naka10,Gomi10,Dayal11} with particular emphasis on
the $\kappa$-phase salts,\cite{remark2} which form a triangular
lattice of dimerized BEDT-TTF molecules that is subject to
frustration.\cite{Mori99,Kandpal09,Powell11,Jeschke12}
 Frequencies of certain intramolecular vibration modes
in BEDT-TTF crystals strongly depend on molecular charge, which
makes optical spectroscopy one of the most sensitive local probes
to investigate the charge
distribution.\cite{Maksimuk01,Dressel04,Yamamoto05,Drichko09,Girlando11,Yakushi12}
We have performed comprehensive in- and out-of-plane  infrared
measurements on \etbr, \etcl{}, and \etcn{}. For none of these
systems the molecular vibrational modes split upon cooling; and
thus there is no indication of charge redistribution as the
temperature is reduced. This result is seemingly at odds with
significant dielectric relaxation found in \etcl{} and \etcn,\cite{Pinteric99,Tomic12}
which has recently been interpreted as a consequence of electronic
ferroelectricity.\cite{AbdelJawad10,Lunkenheimer12}

\section{Materials}
\begin{figure*}
\centering
\includegraphics[trim=12mm 1cm 0cm 21.5cm,clip=true,scale=0.95]{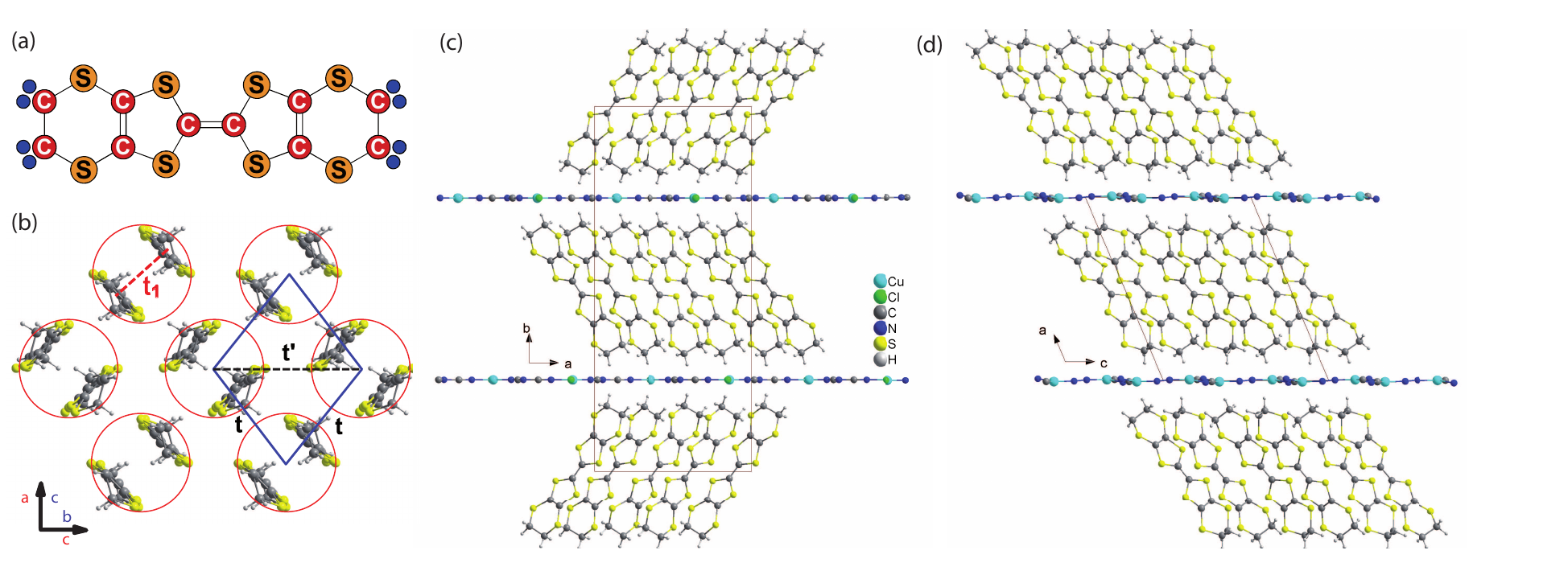}
\caption{\label{fig:1} (Color online) (a) Sketch of the
bis-(ethyl\-ene\-di\-thio)\-te\-tra\-thia\-ful\-va\-lene
(BEDT-TTF) molecule. (b) For $\kappa$-(BEDT-TTF)$_2X$ the
molecules are arranged sideways in dimers (strongly coupled by
$t_1$), which constitute an anisotropic triangular lattice. The
interdimer transfer integrals are labeled by $t$ and $t^{\prime}$.
In the case of \etbr\ and \etcl\ the plane is labeled as $ac$
while it is the $bc$-plane in the case of \etcn, as denoted by the
arrows. (c) The side view on \etcl\ demonstrates the staggered
layers of BEDT-TTF molecules separated by sheets of polymeric
anions. (d) The Structure of \etcn\ consists of one layer tilted
by $\beta=123^{\circ}$ at room temperature ($125^{\circ}$ at low
temperatures).\cite{Jeschke12}}
\end{figure*}
Single crystals of the quasi-two-dimensional organic
charge-transfer salts \etbr{} (abbreviated $\kappa$-Br), \etcl{}
(called $\kappa$-Cl) and \etcn{} ($\kappa$-CN hereafter) were grown
by standard electrochemical methods.\cite{Williams92,Toyota07} In
the crystallographic $\kappa$-phase, conducting layers of
cat\-ionic
bis-(ethyl\-ene\-di\-thio)\-te\-tra\-thia\-ful\-va\-lene
(BEDT-TTF) molecules
are separated by essentially insulating anion
sheets, as depicted in Fig.~\ref{fig:1}. The BEDT-TTF donors form face-to-face dimers
which themselves are rotated by about 90$^{\circ}$ with respect to
neighboring dimers, as sketched in Fig.~\ref{fig:1}(b).

$\kappa$-CN crystallizes in a monoclinic structure (space group
$P2_1/c$) with  the long axis of the BEDT-TTF molecules inclined
by about $25^{\circ}$ with respect to the layer normal,
\cite{Geiser91} as displayed in Fig.~\ref{fig:1}(d). For
convention, in $\kappa$-CN the layer spans in the $bc$-plane and
the $a$-axis points between the layers ($a^*$ is normal to the
$bc$-plane). The Cu$_2$(CN)$_3$-anions form a flat infinite
two-dimensional network of copper and cyano groups in the
$bc$-plane. In the case of isostructural compounds $\kappa$-Br and
$\kappa$-Cl the polymeric anions consist of parallel infinite
zigzag $\cdots$di\-cyano\-amido-Cu(Cl/Br)-di\-cyano\-amido
$\cdots$ chains running along the $a$-direction. The compounds
form orthorhombic (space group $Pnma$) two-layer systems (anion
layers in $ac$-plane, $b$-axis perpendicular to it) with four
(BEDT-TTF)$_2$ dimers per unit cell due to the tilting along $a$
in opposite direction for adjacent layers, as displayed in
Fig.~\ref{fig:1}(c).\cite{Kini90,Williams90}

All three compounds are two-dimensional half-filled electron
systems with considerably strong electronic correlations that
places them close to a metal-insulator transition driven by
effective electronic correlations $U/t$, where $U$ is the on-site
Coulomb repulsion and $t$ the transfer integral. Due to the large
orbital overlap, \kbr{} is the organic superconductor with the
highest transition temperature of 11.8~K at ambient
pressure;\cite{Kini90} upon cooling it crosses over from a bad
metal to a Fermi liquid around $T_{\rm coh} \approx
30$~K.\cite{Yasin11} The sister compound \kcl{} is a Mott
insulator with antiferromagnetic order at low temperatures
($T_N\approx 25$~K),\cite{Welp92,Miyagawa95,remark3} Applying a
small amount of pressure (300 bar) reduces $U/t$ sufficiently to
cross the insulator-metal phase boundary (where it becomes
superconducting at $T_c=12.8$)\cite{Williams90} with unusual
universality class of the critical phenomena at the
pressure-driven Mott transition close to the critical
endpoint.\cite{Kagawa05} Also chemical pressure by Br substitution
is a tuning parameter to investigate the bandwidth-driven Mott
insulator-metal transition.\cite{Yasin11,Faltermeier07} For the
Mott insulator \etcn{} hydrostatic pressure of 1.5~kbar is
required to reach the superconducting state at
$T_c=2.8$~K.\cite{Geiser91,Komatsu96} This compound triggered
particular interest because at ambient pressure no indication of
magnetic order could be observed down to lowest temperatures,
despite the considerable antiferromagnetic exchange of $J\approx
250$~K within the triangular lattice.\cite{Shimizu03}

The transfer integral $t_1$ of the two molecules composing the
dimer  is estimated to be approximately
$0.2$~eV \cite{Komatsu96,Oshima88} and the onsite Coulomb
repulsion $U\approx 2t_1$.\cite{McKenzie98} The coupling between
the dimers is smaller by a factor of 3 to 6 compared to the
intradimer transfer integrals. The effective Coulomb interaction
$U/t$ increases considerably when going from the metall
\kbr{} to the Mott insulator \kcn{}, as summarized in
Tab.~\ref{Tab:1}.
\begin{table}
\caption{Hopping parameters $t^{\prime}$ and $t$ of different
$\kappa$-phase salts obtained from
density functional theory calculations.\cite{Kandpal09} The ratio
$t^{\prime}/t$ is a measure of frustration. In all cases the
intradimer transfer integral is estimated to $t_1=0.2$~eV and the
on-site Coulomb
repulsion $U\approx 2t_1$.\cite{Komatsu96,Oshima88,McKenzie98}
The Hubbard $U$ with respect
to $t$ is a measure of the effective electronic correlations.
\label{Tab:1}}
\begin{center}
\begin{tabular}{c|rr|rrr}
\hline
$\kappa$-(BE\-DT\--TTF)$_2X$ & $t^{\prime}/t$ & $U/t$ & $t_1$ & $t$ & $t^{\prime}$ \\
$X$= &  &  &(meV)&(meV)&(meV)\\
\hline
Cu\-[N\-(CN)$_{2}$]Br  & 0.42  & 5.1 & 200 & 78 & 33 \\
Cu\-[N\-(CN)$_{2}$]Cl & 0.44  & 5.5 & 200 & 73 & 32\\
Cu$_2$(CN)$_{3}$ & 0.83  & 7.3 & 200 & 55 & 45 \\
\hline
\end{tabular}
\end{center}
\end{table}
More important, however, is the degree of
frustration expressed by the ratio of the transfer integrals,
that is $t^{\prime}/t \approx 0.42$ and
0.44 in the case of the Fermi-liquid $\kappa$-Br and the Mott
insulator \kcl, respectively. For the spin-liquid compound
\kcn, the effective Hubbard $U$ is larger ($U/t=7.3$) and
most important the transfer integrals $t^{\prime}/t=0.83$ are very
close to equality.\cite{Kandpal09} Similar values have been
obtained from {\it ab initio} derivations by Nakamura
{\it et al.}\cite{Nakamura09}
For most models and theoretical
descriptions, the interaction between BEDT-TTF layers and anions
is neglected, however, first-principles density-functional theory
calculations recently indicate the importance of donor-anion
hydrogen bonding.\cite{Alemany12}

\begin{figure}
\centering
\includegraphics[trim=0mm 0mm 7cm 18.7cm,clip=true,width=\columnwidth]{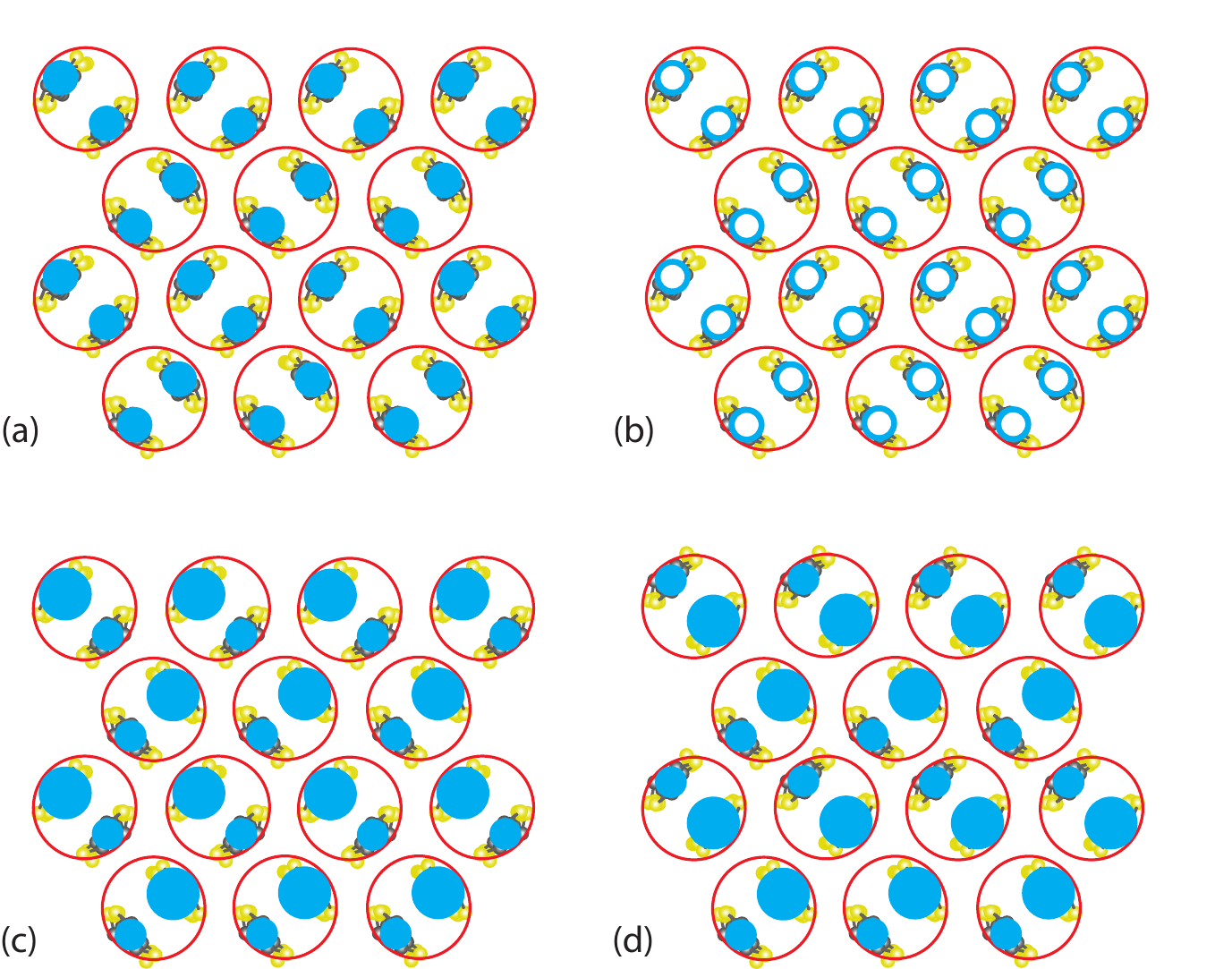}
\caption{\label{fig:2} (Color online) For $\kappa$-(BEDT-TTF)$_2X$
the molecules are arranged in dimers of two BEDT-TTF molecules.
Commonly each molecule is assumed to carry equally charge
(symbolized by the blue dots);  this can be (a) static or (b)
fluctuating. Alternatively, it is suggested that the molecules
carry unequal charge. This charge disproportionation could form a
dipolar solid with the net polarization along (c) vertical or (d)
horizontal direction.}
\end{figure}

Hotta suggested \cite{Hotta10} that despite the strong intradimer
coupling $t_1$ quantum electric dipoles are formed on the dimers
which interact with each other and thus modify the exchange
coupling $J$ between the spins on the dimers, crucial for the
formation of the spin liquid state. Starting with a quarter-filled
extended Hubbard model that includes the interaction $V$ between
molecules on the same but also different dimers, she derives an
effective dipolar-spin model. The quantum electric dipoles
fluctuate by $t_1$. For large $t_1$ a dimer Mott insulator is
stable, forming a dipolar liquid, but if $V$ is large compared to
$t_1$, charge order emerges (dipolar solid), as sketched in Fig.~\ref{fig:2}.
Similar
considerations have been put forward by other groups.\cite{Naka10,Gomi10,Dayal11}

Within the dimer model, Imada and
collaborators \cite{Shinaoka12} presented an {\it ab initio} study
of $\kappa$-phase materials applying a single-band extended
Hubbard model. They discovered that charge fluctuations are
enhanced by the inter-dimer Coulomb interaction.
It is not clear whether the magnetic phase is stable in the
presence of charge fluctuations\cite{Tocchio09} or whether
magnetic and ferroelectric order exclude each other.\cite{Naka10}
Certainly further experiments are needed before
this issue can be resolved.

\section{Vibrational Spectroscopy}
Reflection measurements were performed on as-grown surfaces of
single crystals with light polarized along the two principal
optical axes, that are $E\parallel b$ and $E\parallel c$ in the
case of \kcn\ and $E\parallel c$ and $E\parallel a$ in the case of
\kcl\ and \kbr. Besides regular in-plane experiments, we put
particular emphasis on the polarization perpendicular to the
conducting BEDT-TTF planes, a direction which couples to the most
charge-sensitive infrared-active intramolecular vibrational mode
$\nu_{27}(b_{1\mathrm{u}})$. To this end we utilizing a Bruker
Hyperion infrared microscope attached to the Fourier-transform
spectrometer Bruker IFS 66v/s or Vertex
80v.\cite{Dressel04,Drichko09} The data were collected on the
narrow side of the crystals. While for $\kappa$-CN all three
directions could be well separated, in the case of  $\kappa$-Br
and $\kappa$-Cl the data with $E\parallel b$ and $E\perp b$ might
be a mixture of $a$- and $c$-directions. The sample was cooled
down to $T=12$~K by Cryovac Microstat cold-finger cryostat. For
the in-plane reflection of \kcn\ we also performed
temperature-dependent far-infrared measurements using a Bruker IFS
113v equipped with a cold-finger cyostat and in-situ
gold-evaporation as reference.\cite{Dressel04} In addition the
high-frequency optical properties (up to 35\,000~\cm) were
determined by spectroscopic ellipsometry at room temperature. In
order to perform the Kramer-Kronig analysis using a constant
reflectivity extrapolation at low frequencies and temperatures for
the Mott insulator \kcn\ and \kcl, while a Hagen-Rubens behavior
was assumed for elevated temperatures and for \kbr\ in
general.\cite{DresselGruner02}

In order to follow the vibrational modes down to lower
temperatures, we additionally conducted optical transmission
measurements on a powdered sample. For that purpose single
crystals of \kcn\ were mixed with KBr and ground to fine powder
that was pressed to a 0.35~mm thick disk. The free standing pellet
was placed in a He-bath cryostat and cooled down to $T=2.5$~K and
measured by a Bruker IFS 113v in the mid-infrared range.

\section{Results and Analysis}

\begin{figure}
\centering
\includegraphics[trim=0mm 55mm 0mm 115mm,clip,width=\columnwidth]{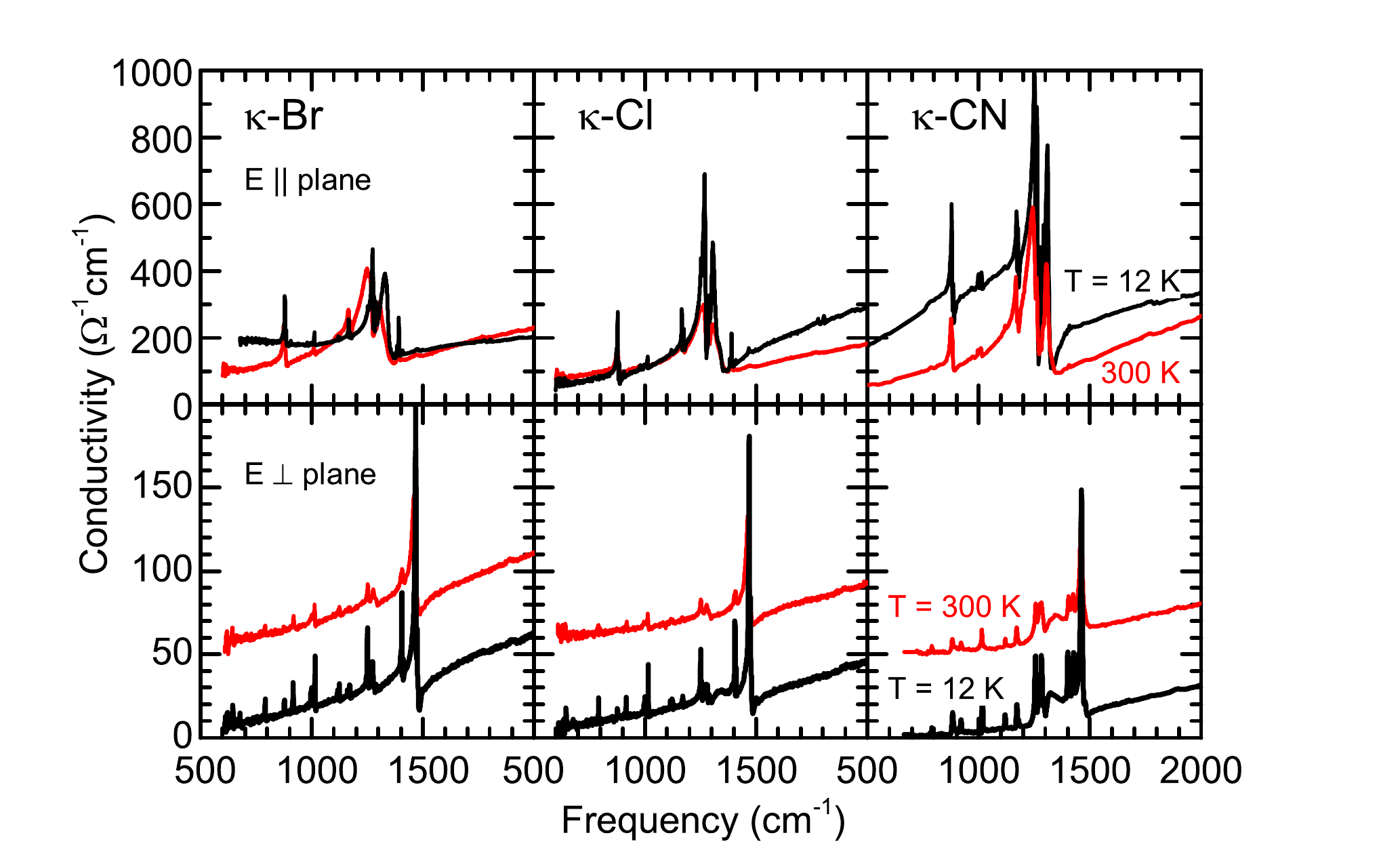}
\caption{\label{fig:3} (Color online) Optical conductivity of
$\kappa$-(BEDT-TTF)$_2X$ measured within (upper frames) and
perpendicular (lower frames) to the BEDT-TTF planes. In the lower
row the room-temperature spectra (red lines) are shifted by
$50~(\Omega{\rm cm})^{-1}$ with respect to the one measured at
$T=12$~K (black curves). For \etbr\ and \etcl, the light was
polarized $E\perp b$ and $E\parallel b$, while for \etcn\ the
polarizations were $E\parallel b$ and $E\parallel a$,
respectively.}
\end{figure}

In Fig.~\ref{fig:3} the frequency-dependent conductivity of \kbr, \kcl\ and \kcn{}
is displayed for the mid-infrared range between 500 and 2000~\cm\
measured at room and low temperatures ($T=12$~K). The in-plane optical spectra (upper panels)
are dominated by the electronic background that changes strongly with temperature,
as investigated previously by several groups.\cite{Eldridge91,Kornelsen92,McGuire01,Kezsmarki06,Faltermeier07,Merino08,Dumm09}
For the out-of-plane conductivity (lower panels) the conductivity is an order of magnitude lower and exhibits no significant temperature dependence in the overall behavior. The vibrational features are well resolved and become narrower upon cooling. The present communication
deals only with the analysis and discussion of the molecular vibrations.

\subsection{In-plane polarization}
\begin{figure}
\centering
\includegraphics[trim=0mm 85mm 0mm 80mm,clip=true,width=\columnwidth]{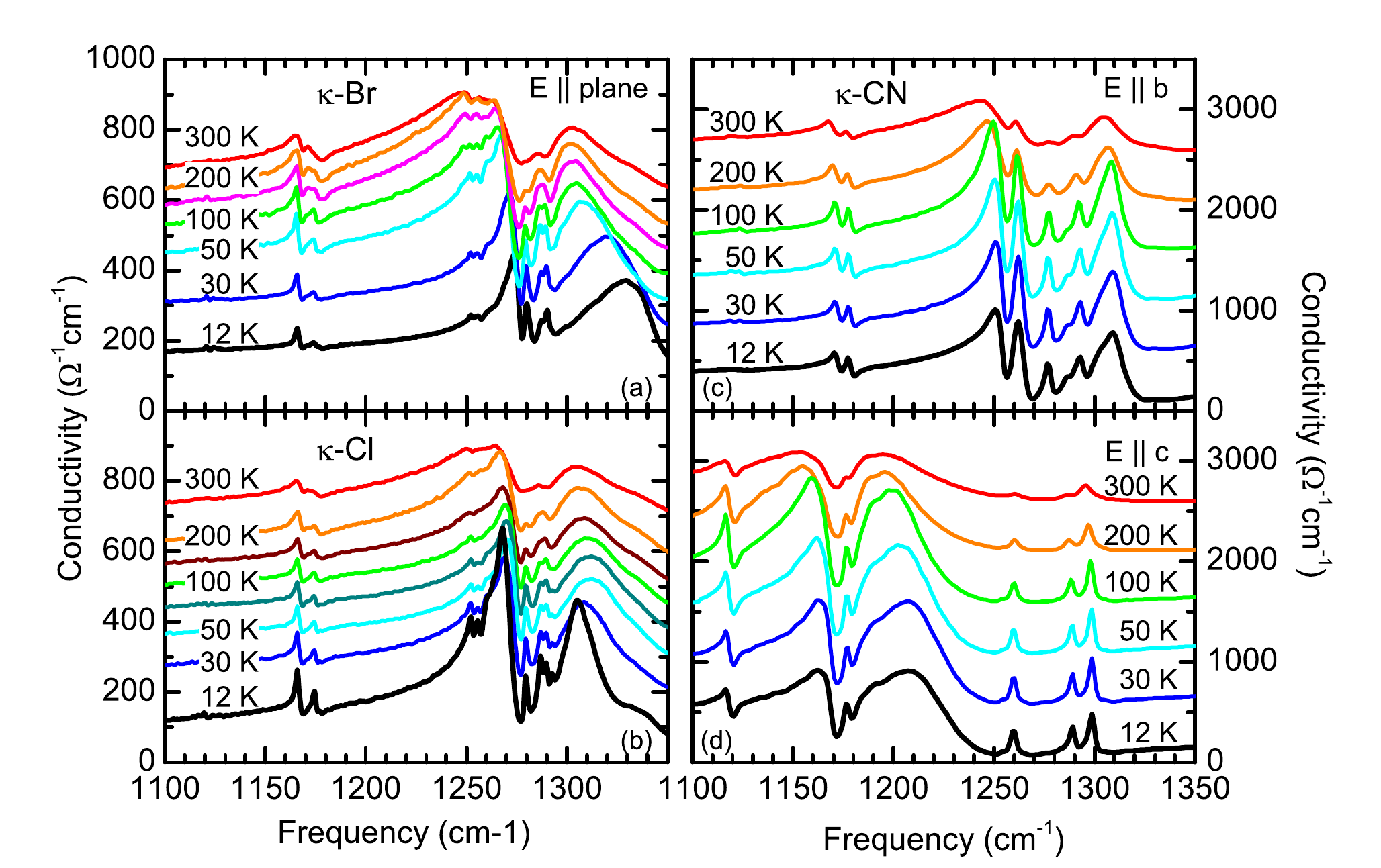}
\caption{\label{fig:4} (Color online) Temperature evolution of the
in-plane optical conductivity of (a) \etbr{} (b) \etcl{}, and
(c,d) \etcn{} in the range of the emv-coupled $\nu_3(a_g)$ mode.
For clarity reasons the curves are shifted with respect to each
other.}
\end{figure}

Expanded portions of the in-plane conductivity spectra are plotted in Fig.~\ref{fig:4} for different temperatures.
The most dominant features near 1300~\cm\ are the totally symmetric $\nu_3(a_g)$ vibrations of the C=C double bonds, which are normally infrared inactive but are activated by electron-molecular vibrational (emv) coupling that involves charge-oscillation between the molecules in each dimer pair. The band is rather broad with a width of approximately 50 to 70~\cm\ that mainly depends on the electronic interaction and thus does not change much with temperature. For the isostructural compounds \kbr\ and \kcl, the feature is basically centered around the same frequency (1275~\cm\ at room temperature).
In $\kappa$-Br the maximum weakens and shifts to higher frequencies as the temperature decreases, indicating a weaker electronic coupling, since the interband conductivity decreases due to the shift of spectral weight to the Drude contribution. The opposite behavior is observed in the Mott insulator \kcl, where the $\nu_3(a_g)$ strength grows dramatically and the peak frequency decreases as the mid-infrared peak grows with the formation of the gap.\cite{Eldridge91,Kornelsen92,Faltermeier07} The four dips between 1377 and 1394~\cm\ belong  to a quartet of $\nu_5(a_g)$  modes and appear as anti\-resonances within the strong $\nu_3(a_g)$ mode.\cite{Kornelsen91}

\begin{figure}
\centering
\includegraphics[trim=0mm 15mm 0mm 15.5cm,clip=true,width=\columnwidth]{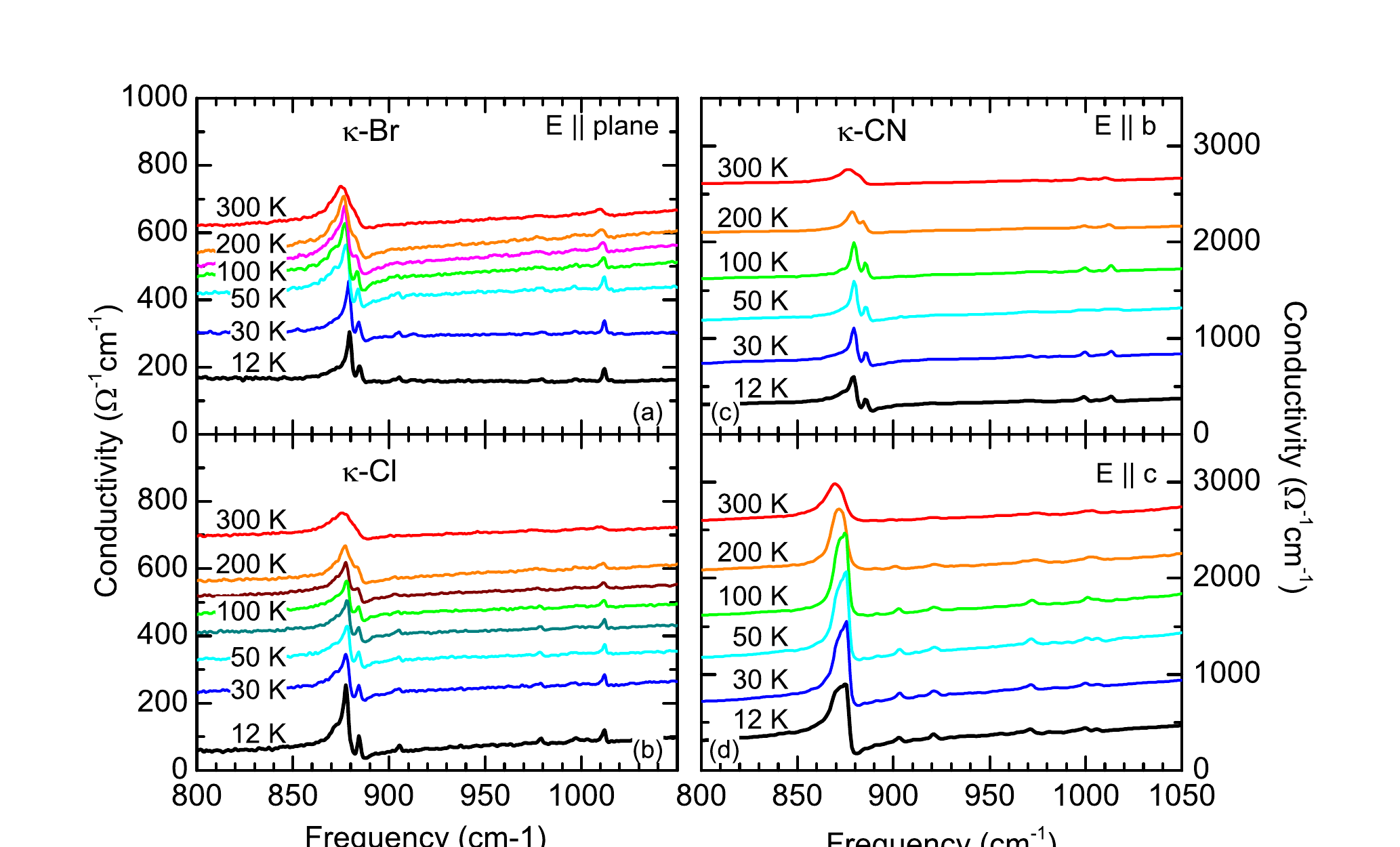}
\caption{\label{fig:5} (Color online) Temperature evolution of the
in-plane optical conductivity of (a) \etbr{} (b) \etcl{}, and
(c,d) \etcn{} in the range of the $\nu_{60}(b_{3u})$ mode. }
\end{figure}

In the case of $\kappa$-CN [Fig.~\ref{fig:4}(c,d)] the vibrational
features are quite different for the two polarization within the
conducting plane. For $E\parallel b$ the emv-coupled $\nu_3(a_g)$
shows up around 1280~\cm, while it is shifted to 1180~\cm\ for
$E\parallel c$, indicating a much stronger emv coupling. This also
agrees with the fact, that the optical properties of \kcn\ are
rather anisotropic compared to \kcl.\cite{Kezsmarki06,Elsasser12}
It is interesting to see, how the dips (anti\-resonances) in one
case become peaks in the other polarization, and vice versa. Note,
the quartet of $\nu_5(a_g)$ dips within the $\nu_3(a_g)$ modes is
degenerated to a doublet for $E\parallel c$. In addition a mode is
seen at 1260~\cm\ that does not split for $E\parallel b$; it is
also present in the other two compounds and can most likely be
assigned to $\nu_{46}(b_{2u})$ or $\nu_{29}(b_{1u})$. In all three
compounds we find a double peak at 1166 and 1174~\cm\ which is
labeled as $\nu_{67}(b_{3u})$. A new Fano-shaped structure occurs
around 1118~\cm\ that might be the $\nu_{47}(b_{2u})$ mode. At
lower frequencies the $\nu_{60}(b_{3u})$ vibrations show up as a
strong doublet at around 880~\cm; see Fig.~\ref{fig:5}. This mode
has drawn considerable attention in the
past\cite{Eldridge98,Lin01,Musfeldt05} and may be regarded as a
totally symmetric mode of a distorted BEDT-TTF molecule. For \kcn\
it converges to a significant Fano shape in the the $E\parallel c$
polarization. This again indicates the anisotropy of \kcn\ and the
particularity of the $c$-direction in this compound.

Most important, in no case do we observe a splitting of any of these modes or some appreciable shift beyond the usual thermal dependence.

\subsection{Out-of-plane polarization}
\begin{figure}[hb]
\centering
\includegraphics[trim=5cm 0mm 5cm 22.5cm,clip=true,width=\columnwidth]{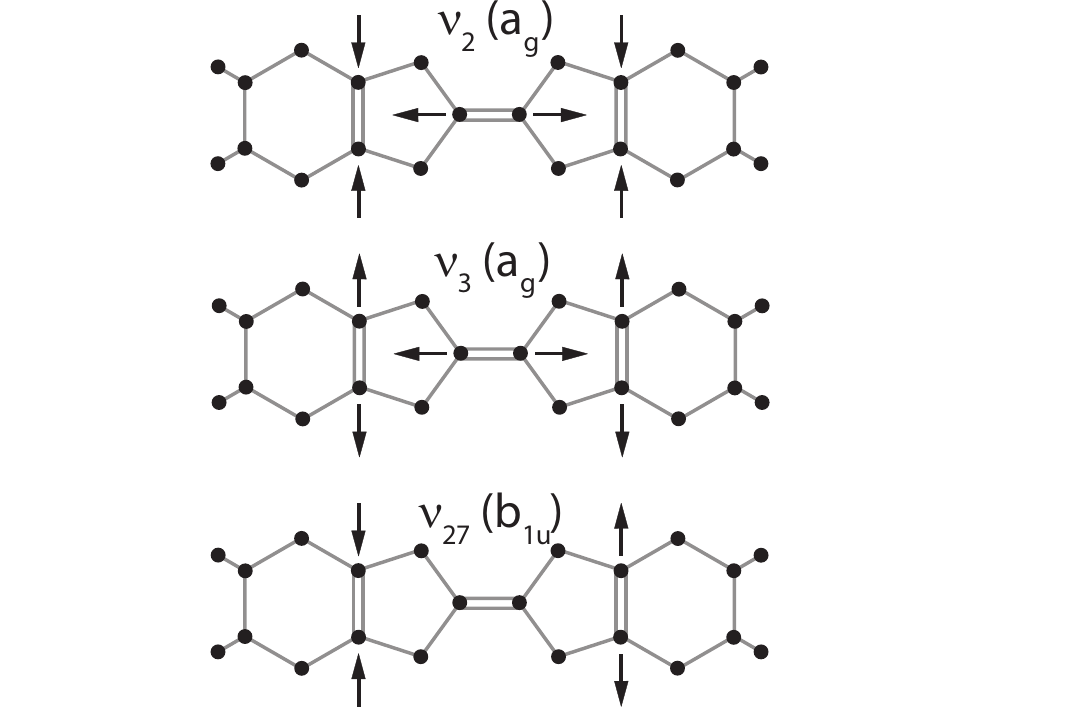}
\caption{\label{fig:6} Sketch of the three most charge-sensitive
C=C stretching modes of the BEDT-TTF molecule: $\nu_2(a_g)$,
$\nu_3(a_g)$, and $\nu_{27}(b_{1u})$, according to $D_{2h}$
symmetry.\cite{Girlando11}}
\end{figure}

As pointed out by Girlando,\cite{Girlando11} Yakushi and col\-labor\-ators \cite{Maksimuk01,Yamamoto05} the totally symmetric molecular vibrations $\nu_2(a_g)$ and $\nu_3(a_g)$ (sketched in Fig.~\ref{fig:6}) exhibit a strong dependence on ionicity of approximately $-125~{\rm cm}^{-1}/e$. Note that the BEDT-TTF molecules are electron donors and thus carry a positive charge of typically 0.5 holes in case of AB$_2$ stoichiometry typically for these compounds.

But these modes are not well suited for infrared determination of the ionicity because they are strongly affected by the electronic coupling strength and location of the mid-infrared band.
The superior local probe of the molecular charge is the antisymmetric stretching mode $\nu_{27}$ that involves the outer  C=C; its eigenfrequency shifts according to\cite{Yamamoto05}
 \begin{equation}
\nu_{27}(\rho)=1398~{\rm cm}^{-1} + 140(1-\rho)~{\rm cm}^{-1} \quad ,
 \end{equation}
where $\rho$ is the site charge. According to the stacking of the BEDT-TTF molecules in the planes, this mode can be best observed  perpendicular to the crystal plane.

\begin{figure}
\centering
\includegraphics[trim=0cm 2.5cm 0cm 3.5cm,clip=true,width=0.8\columnwidth]{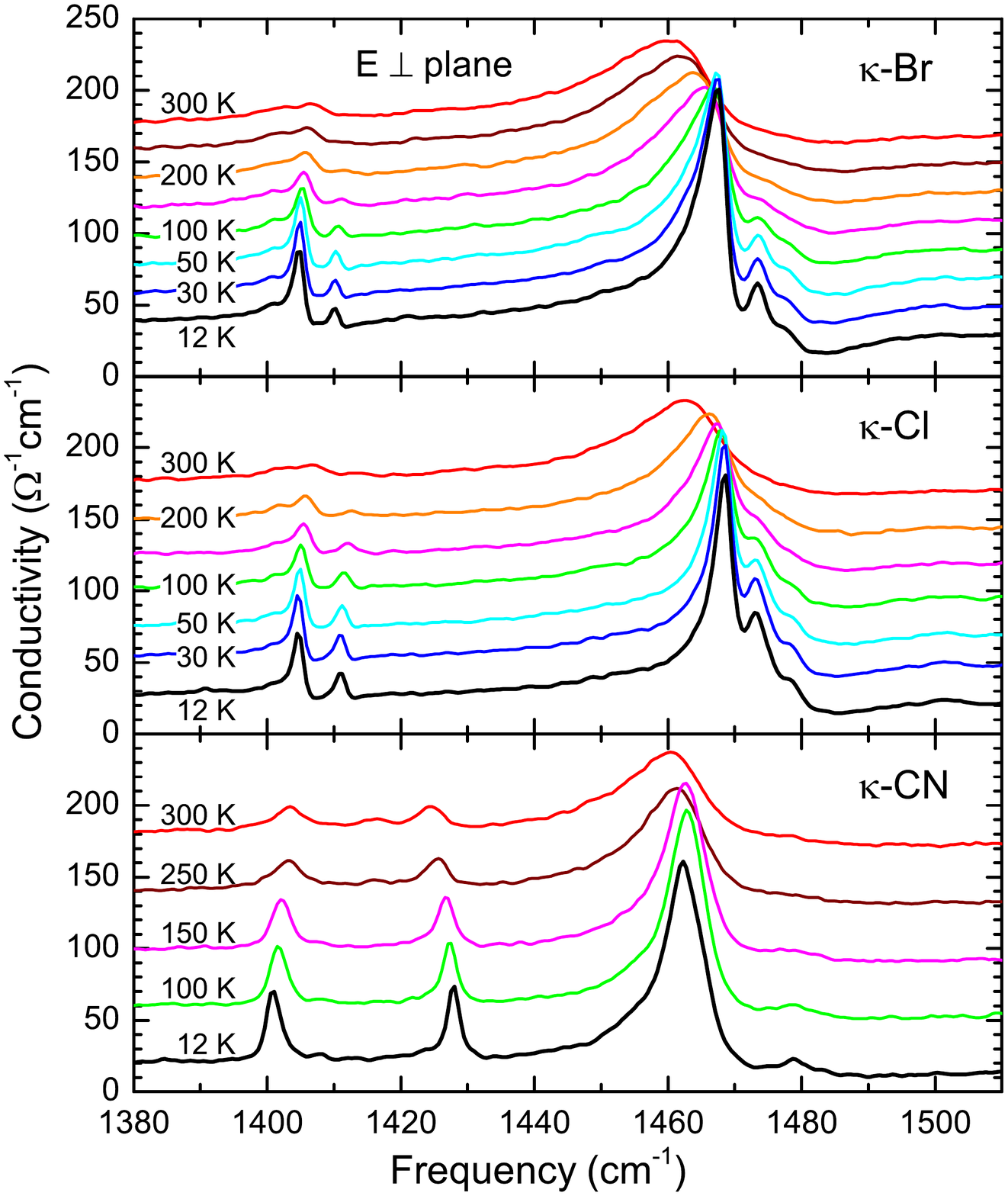}
\caption{\label{fig:7} (Color online) Temperature evolution of the
out-of-plane optical conductivity of \etbr, \etcl{}, and  \etcn{},
measured at the small side of crystals with the electric field
polarized perpendicular to the BEDT-TTF planes. The dominant
vibrational mode $\nu_{27}(b_{1u})$ is a very sensitive local
probe of charge per molecule.}
\end{figure}

In Fig.~\ref{fig:7} the mid-infrared conductivity ($E\perp {\rm
planes}$) of \kbr, \kcl, and \kcn\ is plotted for different
temperatures; at 1460~\cm\ we observe the
$\nu_{27}(b_{1\mathrm{u}})$ mode exactly where it expected for
half a hole per BEDT-TTF molecule. With decreasing temperature
there is a slight hardening of a few \cm\ and a strong narrowing
due to the decrease in interaction with low-lying
phonons.\cite{Maksimuk01} The reduced width of the $\nu_{27}$
feature makes two side bands visible at 1473 and 1478~\cm;
$\kappa$-CN develops a single satellite at 1479~\cm. The reason
lies in the crystallographic inequality among the BEDT-TTF
molecules. \kcn\ consists of a single conducting layer with two
dimers per unit cell, while in \kbr\ and \kcl\ the molecules in
adjacent layers are tilted in opposite directions as depicted in
Fig.~\ref{fig:1}(c). Similarly the $\nu_{28}(b_{1u})$ mode shows
up as a doublet at around 1405 to 1411~\cm\ for the double layer
compounds \kbr\ and \kcl, while as a single line in \kcn; the mode
slightly softens upon cooling. It is interesting to note the mode
at 1428~\cm, which hardens with decreasing temperature, but is
present only in $\kappa$-CN. Temperature-dependent Raman and
infrared experiments on $\kappa$-Br confirm our observations:
Maksimuk {\it et al.}\cite{Maksimuk01} could better describe their
Raman spectra with two Lorentzians at 1468 and
1475~\cm.\cite{remark1}

\subsection{Low-temperature anomaly}
Thermodynamic,\cite{Yamashita08a,Shimizu03}
transport,\cite{Yamashita08b} dielectric \cite{Poirier12} and
lattice \cite{Manna10}  properties provide evidence for a
low-temperature anomaly near $T=6$~K that by now has not been
explained satisfactorily. Lang {\it et al.} suggest that charge
fluctuations are present at elevated temperatures and around 6~K
the preformed electric dipoles exhibit some sort of order.
This might also influence
the magnetic properties. Based on a quarter-filled extended
Hubbard model with both electron-electron and electron-phonon
interaction taken into account, Mazumdar and collaborators
\cite{Dayal11} propose a paired electron crystal that includes
charge-rich and charge-poor molecules.

\begin{figure}[h]
\centering
\includegraphics[trim=0mm 3.5cm 0cm 11.5cm,clip=true,width=0.8\columnwidth]{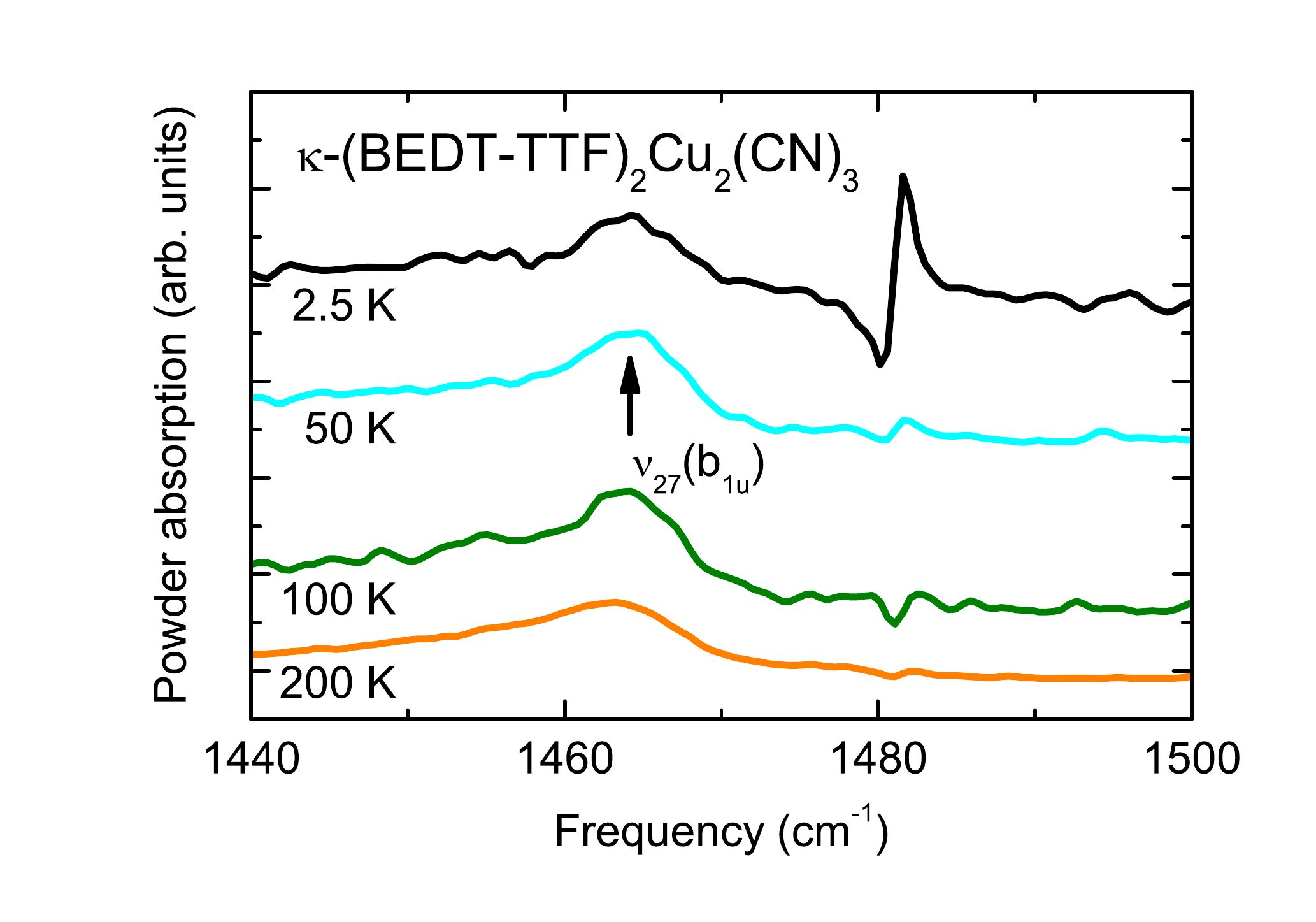}
\caption{\label{fig:8} (Color online) Powder absorption of \etcn\
in the range of the C=C molecular vibrations. The
$\nu_{27}(b_{1u})=1464$~\cm\ mode remains basically unchanged with
temperatures, in accord with the reflection data. When cooling
down to $T=2.5$~K the satellite at 1480~\cm\ becomes more
pronounced and takes a Fano shape. No shift in frequency is
observed as the temperature changes. The curves are displaced for
clarity reasons.}
\end{figure}
In order to experimentally verify whether  a charge rearrangement
happens around the 6~K anomaly, we have performed powder
transmission measurements down to temperatures as low as 2.5~K. In
Fig.~\ref{fig:8} the powder absorption spectra of \kcn\ are
plotted for different temperatures. The peak at 1464~\cm\
corresponds to the $\nu_{27}(a_g)$ mode in accord with the
reflectivity data presented in Fig.~\ref{fig:7}; it slightly
hardens by approximately 1~\cm\ and becomes narrower as the
temperature is reduced. No   significant change is observed when
cooled down to $T=2.5$~K, giving evidence that no charge order
happens at 6~K or any other temperature. Interestingly, the
satellite peak at 1480~\cm\ becomes continuously stronger as the
temperature is reduced from room to helium temperatures; but only
below $T=6$~K it takes a strong Fano shape.\cite{remark5} Dielectric
measurements \cite{AbdelJawad10} give a frequency-independent cusp
near that temperature that indicates some dielectric background to
which molecular vibrations couple, despite the vanishing dc
conductivity. This behavior has been analyzed
elsewhere\cite{Elsasser12} where charge fluctuations close to a
critical point are discussed in more detail.

\section{Discussion}
\subsection{Vibrational features}
In none of the in-plane or out-of-plane spectra displayed above we
could find indications of charge disproportionation, neither in
the metal $\kappa$-Br, the Mott insulator $\kappa$-Cl above or
below the antiferromagnetic ordering, nor in Mott insulator
$\kappa$-CN, which is supposed to be a spin liquid. Since our data
are taken with  1~\cm\ resolution, we can exclude charge imbalance
of more than 3\%. For example, a charge disproportionation of
$2\delta=0.6e$, as typical for charge-ordered systems such as
$\alpha$-(BEDT-TTF)$_2$I$_3$,\cite{Ivek11} results in a splitting
of 80~\cm. Nothing like that is found, neither in our present nor
previous data,\cite{Faltermeier07} nor in temperature- or pressure
dependent Raman and infrared experiments by other
groups.\cite{Truong95,Eldridge95,Eldridge98,Lin01,Maksimuk01,McGuire01,Yamamoto12}
Also complementary methods such as nuclear magnetic resonance
(NMR) or x-ray scattering do not evidence charge
disproportionation in these $\kappa$-phase
compounds.\cite{Shimizu06,remark4} Pictures like
Fig.~\ref{fig:2}(c,d) with a permanent electric dipole can be
ruled out. The experimental findings do not depend on whether the dipoles
are oriented or disordered, they just lead to the conclusion that
there is no charge imbalance. Either the charge degrees of freedom are not involved in the magnetic and thermodynamic phase transition (which is unlikely) or in a more subtle way beyond the simple  arrangements depicted in Fig.~\ref{fig:2}.

When the temperature is reduced below 6~K, several physical
properties exhibit some anomaly indicating a phase transition of
unknown
origin.\cite{Shimizu03,Yamashita08a,Yamashita08b,Manna10,Poirier12}
While the change in the lattice properties implies a modification
of the frustration parameter $t^{\prime}/t$, x-ray investigations
prove that the symmetry does not change, despite the considerable
variations with temperature.\cite{Jeschke12} In particular the
hopping parameter $t^{\prime}$ in $b$-direction  increases below
150~K while $t$ decreases in this temperature range. Furthermore,
the interdimer distance becomes smaller at low temperatures and
thus the coupling $t_1$ increases, except for the lowest
temperature point at $T=5$~K. This is in full agreement with the
conclusions drawn from our optical experiments\cite{Elsasser12}
where we identified a shift of the inter-dimer excitations related
to that. The \kcn\ system is a Mott insulator  due to localization
of one hole per BEDT-TTF dimer and $U$ and $U/t$ become even
stronger when cooled to low temperatures.

In order to more quantitatively analyze the $\nu_{27}(b_{1u})$
mode displayed in Fig.~\ref{fig:7}, we have fitted the vibrational
bands for each temperature. The asymmetry of the features
indicates the interaction of the vibrational excitations with the
electronic background and Lorentz oscillators yield unsatisfactory
fit results. Thus we have utilized the  Fano model
\cite{Fano61,Damascelli97} for which the real and imaginary parts
of the conductivity read
\begin{subequations}
\label{eq:Fano}
\begin{eqnarray}
\sigma_1^{\rm Fano}(\nu) &=&
\sigma_0\frac{\gamma\nu\left[\gamma\nu(q^2-1)+2q(\nu^2-\nu_0^2)\right]}
{(\nu^2-\nu_0^2)^2+\gamma^2\nu^2} \quad ,
\label{eq:Fano1}\\
\sigma_2^{\rm Fano}(\nu) &=&
\sigma_0\frac{\gamma\nu\left[(q^2-1)(\nu^2-\nu_0^2)-2\gamma\nu\right]}
{(\nu^2-\nu_0^2)^2+\gamma^2\nu^2} \quad ,
\label{eq:Fano2}
\end{eqnarray}
\end{subequations}
where $\sigma_0$ is the background and $q$ the phenomenological
coupling parameter. Here strong coupling $q=0$ yields an
anti\-resonance, no coupling $q=\pm\infty$ gives a Lorentzian
shape; the negative sign indicates that the maximum in
$\sigma_1(\nu)$ occurs at low frequencies while the minimum is
located at high frequencies. We define the linewidth
$\gamma=1/2\pi c \tau$, the resonance frequency $\nu_0
=\omega_0/(2\pi c)$, and the spectral weight
$\int\left|\sigma_1(\nu)-\sigma_0\right|\,{\rm d}\nu=\nu_p^2/8$.

\begin{figure}
\centering
\includegraphics[trim=0cm 10cm 0mm 2.8cm,clip=true,width=1\columnwidth]{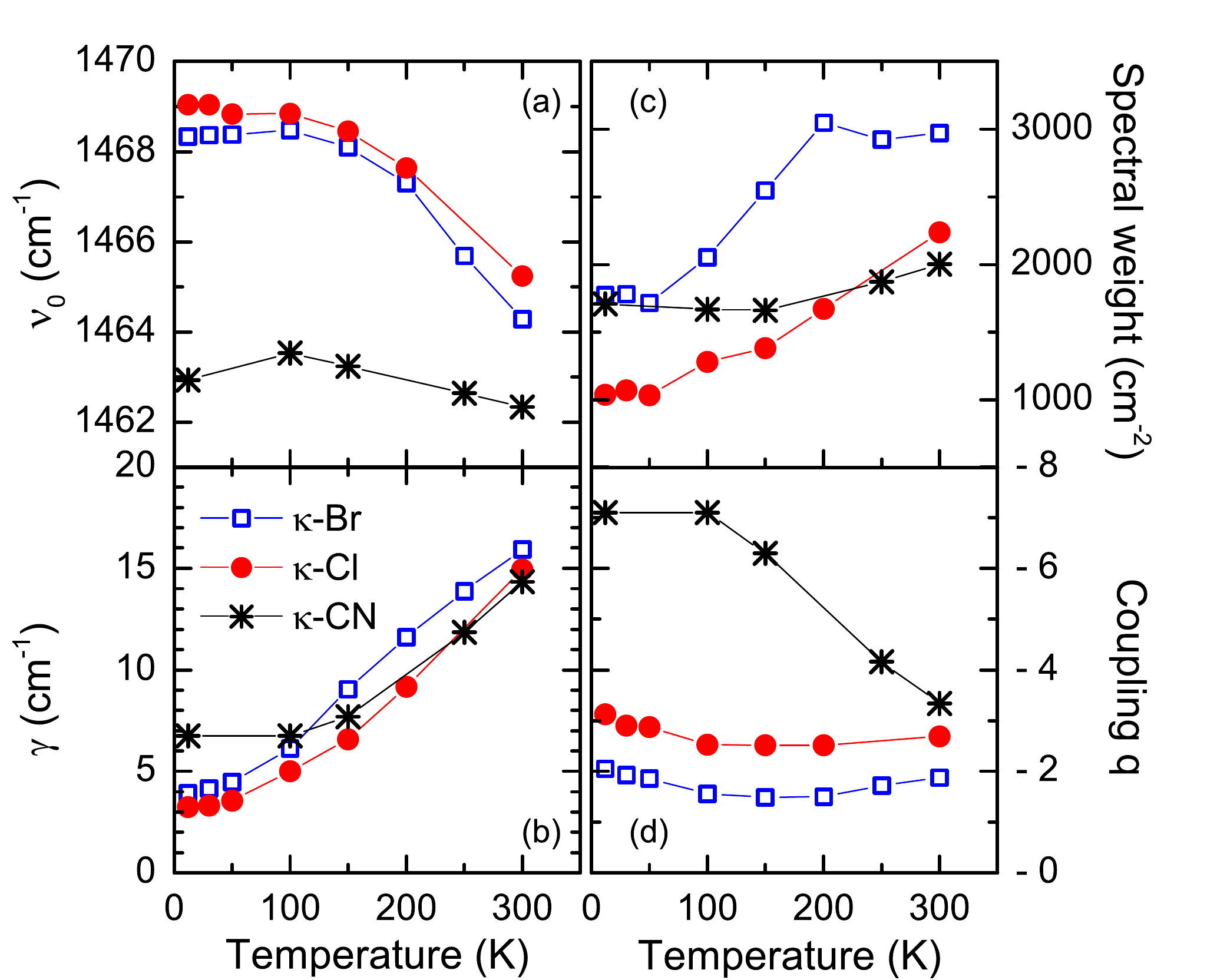}
\caption{\label{fig:9} (Color online)  Temperature dependence of
the mode parameters obtained from the Fano fit of the
$\nu_{27}(a_g)$ vibrational features of \etbr, \etcl, and \etcn.
(a) The resonance frequency $\nu_0$ of the dominant mode, (b) the
damping $\gamma$, (c) the mode strength, and (d) the Fano coupling
parameter $q$.}
\end{figure}
For \kcn\ we observe a small peak around 1480~\cm\ that  calls for
a second Fano term with a spectral weight smaller by a factor of
approximately 300; at very low temperatures ($T\approx 2.5$~K),
the coupling seems to increase strongly, as seen in
Fig.~\ref{fig:8}. In the case of \kbr\ and \kcl\ four modes are
needed, with one significantly stronger, but all of them follow a
very similar temperature dependence.  In Fig.~\ref{fig:9} the mode
parameters of the dominant peaks  for each compound are plotted as
a function of temperature. For all three materials the center
frequency shifts up as the temperature is reduced to approximately
100~K and then basically saturates. There is a slight downshift
observed for \kbr\ and \kcn, but not for \kcl. In all cases the
spectral weight of the mode is significantly reduced when going to
low temperatures [Fig.~\ref{fig:9}(c)]; \kbr\ exhibits some
anomaly around 200~K which cannot be unambiguously assigned at
this point. We are tempted to relate it to the rearrangement of
the ethylene groups that takes place in $\kappa$-(BEDT-TTF)$_2X$
salts at elevated temperatures.\cite{Muller02} Interestingly, for
\kcn\ we observe even a light increase for very low temperatures.
If we consider the peak-to-peak difference $\Delta\sigma$ between
the maximum and minimum in conductivity, for all three compounds
we find a continuous increase from $\Delta\sigma=60~(\Omega{\rm
cm})^{-1}$ to approximately $150~(\Omega{\rm cm})^{-1}$  as the
temperature decreases to 12~K. This is related to the Fano
coupling $q$ and accounts for the asymmetry but also the intensity
of the mode and background conductivity.

\subsection{Fluctuations}
Most noteworthy, the damping gets  considerably smaller upon
cooling with no difference between the metallic or insulating
materials as demonstrated in Fig.~\ref{fig:9}(b); again there
seems to be a saturation below 100~K most obvious in \kcn. The
behavior of the curves is monotonous without any discontinuity;
the magnitude, however, is higher than expected from the typical
narrowing of lattice phonons and molecular vibrations and
indicates that charge fluctuations are present at elevated
temperatures which are reduced when the temperature is lowered.
This behavior is in contrast to the trend commonly observed and
most conventional models. Again for \kcn\ the behavior is slightly
distinct from \kbr\ and \kcl. It is interesting to go back to
Fig.~\ref{fig:3} and compare the conductivity at different
temperatures. While for $E\perp$~planes, there is basically no
temperature change in the mid-infrared conductivity, we do observe
a significant increase for $E\parallel$~plane at low temperatures.
This behavior was investigated in more detail in
Ref.~\onlinecite{Elsasser12}, where we conclude that fluctuations
substantially contribute to the in-plane optical conductivity;
these contributions get enhanced at low temperature. Nevertheless,
charge fluctuations might also be the reason for the strong change
in the damping.

Similar conclusions were drawn by Yamamoto {\it et al.}
\cite{Yamamoto12} from the analysis of the linewidth ($5-10$~\cm)
of the Raman active $\nu_{2}(a_g)$ mode and the infrared active
$\nu_{27}(b_{1u})$ mode and their temperature dependences; they
estimate a charge disproportionation of $2\delta\approx 0.1e$
for the $\kappa$-(BEDT-TTF)$_2X$ salts, which changes only marginally with
temperature. Within the uncertainty of this analysis, these
findings agree with our conclusions.

If we imagine the electronic charge fluctuating within the dimer
and not remaining static as assumed in common models of charge
disproportionation, we can quantitatively estimate the fluctuation
rate. Depending on the time scale of fluctuations and probe, this
results in either a broadening or a splitting of the mode. Kubo
suggested a ``two-states-jump model'' to describe charge to
``jump'' stochastically between two molecules.\cite{Kubo69} Since
the vibrational modes in the charge ordered state do have very
different strength,\cite{Girlando11} the original formula had to
be extended to account for the different infrared intensities,
frequency and band-shape of the mode in differently charges
molecules.\cite{Girlando12} The band-shape function is given by
the real part of the following function:
\begin{equation}
\mathcal{L}(\omega) = \frac{\mathcal{F}[(\gamma+2 v_{\mathrm{ex}})
- {\rm i} (\omega-\omega_\mathrm{w})]} { \mathcal{R}^2-(\omega -
\omega_1)(\omega -\omega_2) - 2 {\rm i}
\Gamma(\omega-\omega_{\mathrm{av}})} \quad . \label{eq:bandshape}
\end{equation}
Here $\mathcal{F} = f_1 + f_2$, with $f_1$, $f_2$ being the
oscillator strengths of the bands of frequency $\omega_1$ and
$\omega_2$ and halfwidth $\gamma$. The charge fluctuation velocity
is $v_{\mathrm{ex}}$, and  $\Gamma = \gamma + v_{\mathrm{ex}}$ the
resulting width, and the abbreviation $\mathcal{R}^2 = 2\gamma
v_{\mathrm{ex}} + \gamma^2$. Finally, we define the average and
weighted frequency, $\omega_{\mathrm{av}}$ and
$\omega_{\mathrm{w}}$, by:
\begin{equation}
\omega_{\mathrm{av}}=\frac{\omega_1+\omega_2}{2};\hspace*{10mm}
\omega_{\mathrm{w}} = \frac{f_2 \omega_1 + f_1 \omega_2}{f_1 +
f_2}. \label{eq:omega_av}
\end{equation}
If the charge oscillations are slow, $v_{\mathrm{ex}} \ll
|\omega_1 - \omega_2|/2$, Eq.~(\ref{eq:bandshape}) yields two
separated bands centered around $\omega_1$ and $\omega_2$, while
if $v_{\mathrm{ex}} \gg |\omega_1 - \omega_2|/2$, the motional
narrowing will give one single band centred at the intermediate
frequency $\omega_{\mathrm{av}}$. Finally, when $v_{\mathrm{ex}}
\approx |\omega_1 - \omega_2|/2$ we shall observe one broad band
shifted towards the mode with higher oscillator strength.
Since we do not see any splitting of the $\nu_{27}(b_{1u})$ mode
and the peaks become narrower as the temperature is reduced,
instead of broader as expected from an growing charge
disproportionation, the charges are either static or fluctuate
with a rate of $10^{14}$~Hz or more which bring us
into the range of electronic transitions.

Another interesting point previously made by Yakushi and
collaborators\cite{Maksimuk01} is the eightfold (fourfold)
splitting of every internal molecular vibration due to the large
unit cell of \kbr\ and \kcl\ (in the case of \kcn). Considering a
dimer inside a crystal, these components are divided into $a_g$
and $a_u$, since the two molecules in a dimer are connected by a
center of symmetry. In the case of {\em gerade} molecular modes,
the molecules in a dimer oscillate in phase or out of phase,
respectively. The interesting point is, that while the $A_g$ and
$B_{1g}$ modes and the $B_{2g}$ and $B_{3g}$ modes in the crystal
are nearly degenerate due to the weak interlayer interaction, the
degeneracy of the $A_{u}$ and $B_{1u}$, as well as between
$A_{2u}$ and $B_{3u}$ is likely to be lifted due to dipole-dipole
interaction between the dipoles induced within the layers. This
effects accounts for the splitting we observe on various modes in
our in-plane and out-of-plane measurements.

It is worthwhile to note, that the electronic coupling between
vibrational excitations, either within the dimer or between
dimers, happens instantaneously compared to the time scale of the
atomic motion (Frank-Condon principle). There is no charge
transfer involved that could affect the molecular vibrations
nor influence the dielectric properties happening in the
radio-frequency range and even lower.

\subsection{Anisotropy}
Let us come back to the remarkable anisotropy observed for \kcn,
as demonstrated in Fig.~\ref{fig:4}(c,d). There seems to be a much
stronger coupling of the $\nu_3(a_g)$ vibrational modes in
$c$-direction compared to the $b$-direction, resulting in a larger
downshift due to emv interaction. Since this is the plane the
BEDT-TTF molecules are tilted in, it might be related to the
coupling of the molecular vibrations to the in-plane excitations.
As the temperature is lowered the tilting even increases by
approximately $1^{\circ}$ which enhances this
effect.\cite{Jeschke12} The major difference, that exists to the
\kbr\ and \kcl\ properties, with one of them conducting the other
insulating, however, implies that simple emv coupling to the
conducting electrons is not solely responsible. More likely it is
the coupling to the anions that form a two-dimensional network in
the $bc$-plane. From x-ray diffraction results,\cite{Jeschke12} it
is known that with decreasing temperatures the Cu$_2$(CN)$_3$
network becomes warped possibly due to the unbalanced interaction
with the ethylene groups. A stronger coupling of the anions to one
end of the BEDT-TTF molecules also explains why totally symmetric
molecular vibrations become infrared active, as in the case of the
$\nu_6(a_g)$ intramolecular vibration. This infers a loss of
symmetry (loss of inversion centers) although the translation
symmetry is still preserved. It also is consistent with the
intriguing results of thermal expansion, that evidence an
anomalous behavior around 6~K. From this point of view it is
interesting to note, that the dielectric response reported by
Abdel-Jawad {\it et al.} was obtained from experiments
perpendicular to the molecular layers.\cite{AbdelJawad10} The
dielectric constant was found to be rather small ($\epsilon_1
\approx 10$) compared to \kcl\ ($\epsilon_1 \approx
400$).\cite{Lunkenheimer12} The interaction of positively and
negatively charged planes might be the crucial point and requires
further considerations. It is of paramount interest to explore the
the anisotropy we predict for the dielectric properties within the
$bc$ plane of the triangular lattice of \etcn.

\section{Conclusions}
We have performed comprehensive measurements of the vibrational
features of the dimerized $\kappa$-(BEDT-TTF)$_2X$ salts, ranging
from the superconducting metal \etbr\ to the Mott insulator \etcl\
and the spin-liquid compound \etcn. Despite the large differences
in electronic and magnetic properties, the vibrational features
are remarkably similar. In particular focussing on the
antisymmetric $\nu_{27}(b_{1u})$ mode observed in out-of-plane
measurements, we can rule out any charge disproportionation within
the dimers, and charge fluctuations of a moderate rate. No
modification in the charge distribution is observed at any temperature;
this also holds for the low-temperature anomaly
reported for the spin-liquid $\kappa$-(BE\-DT\--TTF)$_2$\-Cu$_2$(CN)$_{3}$ around $T=6$~K. This implies that
the dielectric relaxation
observed in the insulating samples \etcl\ and \etcn\ is not due to the formation of
paired electron crystals or quantum electric dipoles. We suggest
to reconsider the interaction of the molecular layers to the
polymeric anion sheet.

\acknowledgements We thank N. Drichko, A. Girlando, K. Kanoda,
J.-P. Pouget and T. Yamamoto for helpful discussions. D.W.
acknowledges support by the Alexander von Humboldt foundation. The
project was supported by the Deutsche Forschungsgemeinschaft (DFG)
and by UChicago Argonne, LLC, Operator of Argonne National
Laboratory ("Argonne"). Argonne, a U.S. Department of Energy
Office of Science laboratory, is operated under Contract No.
DE-AC02-06CH11357. We also acknowledge support by the Croatian
Ministry of Science, Education and Sports under Grant
035-0000000-2836.


\begin{thebibliography}{99}
\bibitem{remark1}
It is interesting to note that the splitting did no occur in
deuterated crystals, indicating the importance of the coupling to
the anionic layers by the methyl groups.
\bibitem{remark5}
A similar Fano-shaped resonance at 987~\cm\ becomes very
pronounced as the temperature is reduced to $T=2.5$~K. We assign
it to the $\nu_6(a_g)$ mode that is rather strong in the Raman
response but has been observed in infrared spectra previously
(Ref.~\onlinecite{Eldridge95}). It mainly involved the outer ring
and the terminating ethylene groups; implying that the mode is
affected by the asymmetric coupling to the anion layers.
\end{thebibliography}
\end{document}